\documentclass[final,5p,times,twocolumn]{elsarticle}

\usepackage{amsmath,amsfonts,amssymb,bm}
\usepackage{color}
\usepackage{dsfont}
\usepackage{graphicx}
\usepackage{multirow}
\usepackage{soul}

\usepackage[mathscr,scaled=1.15]{urwchancal}
\DeclareFontFamily{OT1}{pzc}{}
\DeclareFontShape{OT1}{pzc}{m}{it}%
{<-> s * [1.15] pzcmi7t}{}
\DeclareMathAlphabet{\mathpzc}{OT1}{pzc}{m}{it}

\definecolor{purple}{rgb}{0.5,0,0.5}
\definecolor{blue}{rgb}{0.0,0,0.9}
\definecolor{prdblue}{rgb}{0.133,0.118,0.498}
\usepackage[colorlinks=true, pdfstartview=FitV, linkcolor=prdblue, citecolor= prdblue, urlcolor=prdblue]{hyperref}

\biboptions{sort&compress}

\journal{Physics Letters B}

\begin{document}

\begin{frontmatter}

\title{Distribution Amplitudes of Heavy-Light Mesons}

\author[ECT]{Daniele Binosi}
\author[NKU]{Lei Chang}
%\email[]{leichang@nankai.edu.cn}
\author[ECT,NKU,ANL]{Minghui Ding}
\author[UV]{Fei Gao}
\author[UV]{Joannis Papavassiliou}
%\email[]{joannis.papavassiliou@uv.es}
\author[ANL]{Craig D. Roberts}%s\corref{cor2}}
%\email[]{cdroberts@anl.gov}

\address[ECT]{European Centre for Theoretical Studies in Nuclear Physics
and Related Areas (ECT$^\ast$) and Fondazione Bruno Kessler\\ Villa Tambosi, Strada delle Tabarelle 286, I-38123 Villazzano (TN) Italy}
\address[NKU]{
School of Physics, Nankai University, Tianjin 300071, China}
\address[ANL]{
Physics Division, Argonne National Laboratory, Lemont, Illinois
60439, USA}
\address[UV]{
Department of Theoretical Physics and IFIC, University of Valencia and CSIC, E-46100, Valencia, Spain}

%\date{03 December 2018}
%\date{24 November 2018}

%\cortext[cor1]{Corresponding author: leichang@nankai.edu.cn}
%\cortext[cor2]{Corresponding author: cdroberts@anl.gov}

\begin{abstract}
A symmetry-preserving approach to the continuum bound-state problem in quantum field theory is used to calculate the masses, leptonic decay constants and light-front distribution amplitudes of empirically accessible heavy-light mesons.
The inverse moment of the $B$-meson distribution is particularly important in treatments of exclusive $B$-decays using effective field theory and the factorisation formalism; and its value is therefore computed: $\lambda_B(\zeta = 2\,{\rm GeV}) = 0.54(3)\,$GeV.
As an example and in anticipation of precision measurements at new-generation $B$-factories, the branching fraction for the rare $B\to \gamma(E_\gamma) \ell \nu_\ell$ radiative decay is also calculated, retaining $1/m_B^2$ and $1/E_\gamma^2$ corrections to the differential decay width, with the result $\Gamma_{B\to \gamma \ell \nu_\ell}/\Gamma_B = 0.47(15)$ on $E_\gamma > 1.5\,$GeV.
\end{abstract}


\end{frontmatter}

\setcounter{section}{1}

\noindent\textbf{\arabic{section}.\,Introduction}\,---\,In quantum chromodynamics (QCD), numerous hard exclusive processes can be analysed using the factorisation formalism.  Prominent examples are the applications to elastic and transition form factors of pseudoscalar mesons \cite{Lepage:1979zb, Efremov:1979qk, Lepage:1980fj}.  Such treatments separate the amplitude for a given scattering process into short- and long-distance components: the short-distance part is calculable using perturbative QCD; but the long-distance piece is essentially nonperturbative, deriving from the wave function of the participating hadron.  It was early appreciated that factorisation can also be employed in the treatment of exclusive decays of heavy mesons \cite{Szczepaniak:1990dt}; and the framework has subsequently been cleanly defined and widely employed  -- see, \emph{e.g}.\ Refs.\,\cite{Beneke:1999br, Beneke:2000ry, Beneke:2001at} and citations thereof.

Constituted from a valence $\bar b$-quark and either a valence $u$- or $d$-quark, the $B^{(+,0)}$ are the most widely studied heavy mesons, with high-profile measurements completed, underway, and planned \cite{Bevan:2014iga, Adeva:2017zlz, Kou:2018nap}.
% Past = Belle ... BaBar
% Now = LHCB
% Future = Belle II
Numerous exclusive processes involving this system are well suited to treatment via the factorisation approach. Each associated formula features $\varphi_B$, the $B$-meson light-front distribution amplitude (DA), which is a direct analogue of the light-meson DAs that appear in the earliest factorisation formulae.  However, whilst much has recently been learnt about the pointwise behaviour of leading-twist light-meson DAs \cite{Chang:2013pq, Chang:2013nia, Segovia:2013eca, Gao:2014bca, Braun:2015axa, Shi:2015esa, Raya:2015gva, Li:2016dzv, Li:2016mah, Gao:2016jka, Chang:2016ouf, Zhang:2017bzy, Gao:2017mmp, Chen:2017gck, Chen:2018rwz, Ding:2018xwy}, information about the $B$-meson DA remains sketchy \cite{Braun:2003wx, Lee:2005gza, Grozin:2005iz, Pilipp:2007sb, Beneke:2011nf, Braun:2012kp, Bell:2013tfa, Beneke:2018wjp}.

Considered as a function of $\xi$, the light-front longitudinal momentum fraction of the light-quark in the $B$-meson, it is known that at resolving scales, $\zeta$, very much in excess of the $B$-meson mass, $m_B$, $\varphi_B(\xi) \approx 6 \xi (1-\xi)$.  On the other hand, on $\zeta \lesssim m_B$, $\varphi_B(\xi)$ must be a very asymmetric function, whose peak lies at $\xi \simeq \hat w / m_B$, where $\hat w >0$ is an intrinsic momentum-scale associated with the dressed light-quark in the $B$-meson.  More information is required, however, before factorised formulae for exclusive processes involving $B$-mesons can be useful.  Herein, therefore, we will employ a continuum approach to the hadron bound-state problem in order to compute the pointwise behaviour of the $B$-meson DA at a typical hadronic scale, omitting radiative corrections \cite{Braun:2003wx, Lee:2005gza}; the DAs of other heavy-light systems; and an array of derived quantities, including the branching fraction for the $B\to \gamma \ell \nu_\ell$ radiative decay.
%% omitting radiative corrections means at that point where the regular and HQET PDA are the same.

\smallskip

\addtocounter{section}{1}

\noindent\textbf{\arabic{section}.\,Distribution Amplitudes}\,---\,Consider a heavy pseudoscalar meson with mass $M_h$ and total momentum $p=M_h v$, $v^2=-1$, constituted from a single heavy valence $\bar Q$-quark and a lighter $l$-quark; then one may define a distribution amplitude for this system as the following light-front projection of the meson's Poincar\'e-covariant Bethe-Salpeter wave function:
\begin{subequations}
\label{HMDA}
\begin{align}
f_{h} M_h \tilde\varphi_h(w;\zeta) & =
{\rm tr}_{\rm CD} Z_2 \int_{dk}^\Lambda \delta(n\cdot k - w)
\gamma_5 \gamma\cdot n \chi_h(k;p)\,,\\
\chi_h(k;p) & = S_l(k) \Gamma_h(k;p) S_{Q}(k-p)\,.
\end{align}
\end{subequations}
Here: $f_h$ is the meson's leptonic decay constant; the trace is over colour and spinor indices; $\int_{dk}^\Lambda$ is a Poincar\'e-invariant regularization of the four-dimensional integral, with $\Lambda$ the ultraviolet regularization mass-scale;
$Z_{2}(\zeta,\Lambda)$ is the mass-independent quark wave-function renormalisation constant \cite{Chang:2008ec}, with $\zeta$ the renormalisation scale;
$n$ is a light-like four-vector, $n^2=0$, $n\cdot v = 1$;
$w = \xi n\cdot p$;
$S_{l,Q}$ are dressed-propagators for the meson's valence quarks; and $\Gamma_h(k;p)$ is the meson's Bethe-Salpeter amplitude.  It can be shown \cite{Ivanov:1997yg, Ivanov:1998ms} that in the limit $M_h \to \infty$, $\Gamma_h(k;p) \propto  \hat\Gamma_h(k;v) \surd M_h$ and, \emph{e.g}.\ $f_h \surd M_h =\,$constant.

The DA defined in Eqs.\,\eqref{HMDA} has mass-dimension negative-one, support on $w\in[0,M_h]$, and is unit normalised.  It follows that one can define
\begin{equation}
\label{phitildephi}
\varphi_h(\xi;\zeta) = M_h \tilde\varphi_h(M_h \xi ;\zeta)\,,\quad
\int_0^{1} d\xi \, \varphi_h(\xi;\zeta)  = 1\,.
\end{equation}
QCD-evolution on $\zeta \lesssim M_h$ actually extends the domain of support to $w\in[0,\infty)$ by introducing a radiative tail \cite{Beneke:2000ry}.  We avoid this issue herein by computing all results at a low hadronic scale $\zeta = \zeta_2 = 2\,$GeV, from which evolution can subsequently be employed, if desired.

\smallskip

\addtocounter{section}{1}

\noindent\textbf{\arabic{section}.\,Bound-State Problem}\,---\,Our calculation of $\varphi_h(\xi;\zeta)$ proceeds as follows.
(\emph{i}) Specify a symmetry-preserving truncation of the continuum bound-state problem.
(\emph{ii}) Using that truncation, compute the dressed-quark propagators and meson Bethe-Salpeter amplitude.
(\emph{iii}) Evaluate the DA by inserting the results in Eqs.\,\eqref{HMDA}, \eqref{phitildephi}.
We now elaborate on each of these steps.

The continuum bound-state problem is defined by a set of coupled integral equations \cite{Roberts:2015lja, Eichmann:2016yit}.  A tractable system is only obtained once a truncation scheme is specified.  A systematic, symmetry-preserving approach is described in Refs.\,\cite{Munczek:1994zz, Bender:1996bb}.  The leading-order term is the widely-used rainbow-ladder (RL) truncation.  It is accurate for ground-state light-quark vector- and isospin-nonzero-pseudoscalar-mesons, and related ground-state octet and decouplet baryons \cite{Chang:2011vu, Bashir:2012fs, Roberts:2015lja, Horn:2016rip, Eichmann:2016yit}; and, with judicious modification, heavy-heavy $S$-wave quarkonia \cite{Ding:2015rkn}.  RL truncation is accurate in these channels because corrections largely cancel owing to preservation of relevant Ward-Green-Takahashi identities \cite{Ward:1950xp, Green:1953te, Takahashi:1957xn} ensured by the scheme \cite{Munczek:1994zz, Bender:1996bb}.

On the other hand, in systems constituted from valence-quarks with different renormalisation group invariant (RGI) current-masses: $\hat\delta_{Qq} = \hat m_Q-\hat m_q$, there is typically a maximum acceptable value of this difference, $\hat\delta_{Qq}^{\rm cr}$, such that RL truncation becomes a poor approximation on $\hat\delta_{Qq} > \hat\delta_{Qq}^{\rm cr}$, because the disparity in masses is then too large for the cancellation of corrections to be effective.\footnote{There is a correlated issue: owing to moving singularities in the complex-$k^2$ domain sampled by the bound-state equations \cite{Maris:1997tm}, it can become difficult in practice to obtain a reliable solution on $\hat\delta_{Qq} > \hat\delta_{Qq}^{{\rm cr}^\prime}$.  The value of the ratio $\hat\delta_{Qq}^{\rm cr}/\hat\delta_{Qq}^{{\rm cr}^\prime}$ depends on $\hat m_Q$.
}
Truncations which improve upon RL are known \cite{Chang:2009zb, Bashir:2011dp, Williams:2015cvx, Binosi:2016rxz, Qin:2016fbu, Binosi:2016wcx}, but they have not been tested in heavy-light systems.  We therefore use RL truncation on $\hat\delta_{Qq} < \hat\delta_{Qq}^{\rm cr}$; and extrapolate all computed quantities into the complementary domain using the Schlessinger point method (SPM), whose properties and accuracy are explained elsewhere \cite{Schlessinger:1966zz, PhysRev.167.1411, Tripolt:2016cya, Chen:2018nsg}.

An efficacious RL kernel for the gap and Bethe-Salpeter equations is detailed in Refs.\,\cite{Qin:2011dd, Qin:2011xq}:
 \begin{subequations}
\label{KDinteraction}
\begin{align}
\mathscr{K}_{\alpha_1\alpha_1',\alpha_2\alpha_2'} & = {\mathpzc G}_{\mu\nu}(k) [i\gamma_\mu]_{\alpha_1\alpha_1'} [i\gamma_\nu]_{\alpha_2\alpha_2'}\,,\\
 {\mathpzc G}_{\mu\nu}(k) & = \tilde{\mathpzc G}(k^2) T_{\mu\nu}(k)\,,
\end{align}
\end{subequations}
with $k^2 T_{\mu\nu}(k) = k^2 \delta_{\mu\nu} - k_\mu k_\nu$ and ($s=k^2$)
\begin{align}
\label{defcalG}
 \tfrac{1}{Z_2^2}\tilde{\mathpzc G}(s) & =
 \frac{8\pi^2}{\omega^4} D e^{-s/\omega^2} + \frac{8\pi^2 \gamma_m \mathcal{F}(s)}{\ln\big[ \tau+(1+s/\Lambda_{\rm QCD}^2)^2 \big]}\,,
\end{align}
%KDinteractiondefcalG
where $\gamma_m=4/\beta_0$, $\beta_0 = 11 - (2/3) n_f$, $n_f=4$, $\Lambda_{\rm QCD}=0.234\,$GeV, $\tau={\rm e}^2-1$, and ${\cal F}(s) = \{1 - \exp(-s/[4 m_t^2])\}/s$, $m_t=0.5\,$GeV.

The development of Eqs.\,\eqref{KDinteraction}, \eqref{defcalG} is summarised in Ref.\,\cite{Qin:2011dd} and their connection with QCD is described in Ref.\,\cite{Binosi:2014aea}.  The kernel seemingly depends on two parameters.  However, in baryons and mesons formed from heavy quarks, many observable properties are practically insensitive to variations of $\omega \in [0.7,0.9]\,$GeV, so long as
%\begin{equation}
$\varsigma^3 := D\omega = {\rm constant}$ \cite{Chen:2016bpj, Qin:2018dqp},
%\end{equation}
with empirical values reproduced using
\begin{equation}
\label{Dwconstant}
\varsigma = 0.6\,{\rm GeV}.
\end{equation}
Herein, we employ $\omega = 0.8\,$GeV, the midpoint of the insensitivity domain.  With these values one obtains a kernel in agreement with the RGI interaction derived from analyses of QCD's gauge sector \cite{Binosi:2014aea, Binosi:2016nme, Rodriguez-Quintero:2018wma}.

With the kernel now specified, we perform a coupled solution of the dressed-quark gap-  and meson Bethe-Salpeter-equations, varying the gap equations' current-quark masses until the Bethe-Salpeter equation has a solution at $P^2= - M_{h}^2$, following Ref.\,\cite{Maris:1997tm} and adapting the algorithm improvements from Ref.\,\cite{Krassnigg:2009gd} when necessary.  The benchmarking results in Table~\ref{TableBench} were obtained using RGI current-masses $\hat m_b = 7.4\,$GeV, $\hat m_c = 1.7\,$GeV.  They correspond to the following values of the dressed-quark mass-functions:
\begin{equation}
\label{Mhzeta2}
m_b:= M_b(\zeta_2) = 4.35\,{\rm GeV}\,,\;  m_c:= M_c(\zeta_2) = 1.25\,{\rm GeV}\,,
\end{equation}
defining current-quark masses which are commensurate with other determinations \cite{Tanabashi:2018oca}.

\begin{table}[t]
\caption{\label{TableBench}
Masses and decay constants of heavy pseudscalar mesons ($\eta_c$, $\eta_b$) computed herein, using Eqs.\,\eqref{KDinteraction}\,--\,\eqref{Mhzeta2}; compared with available experimental \cite{Tanabashi:2018oca} and lattice-QCD determinations \cite{Davies:2010ip, McNeile:2012qf}.
(Quantities listed in GeV.)\vspace*{0.5ex}
}
\begin{tabular*}%{l||*{2}{c|}|*{2}{c|}|*{2}{c|}}
{\hsize}
{l@{\extracolsep{0ptplus1fil}}
|c@{\extracolsep{0ptplus1fil}}
c@{\extracolsep{0ptplus1fil}}
|c@{\extracolsep{0ptplus1fil}}
c@{\extracolsep{0ptplus1fil}}
|c@{\extracolsep{0ptplus1fil}}
c@{\extracolsep{0ptplus1fil}}}
\hline
\multicolumn{1}{l}{} & \multicolumn{2}{c}{ Herein } & \multicolumn{2}{c}{ Exp.\,\mbox{\cite{Tanabashi:2018oca}} }& \multicolumn{2}{c}{ lQCD \mbox{\cite{Davies:2010ip, McNeile:2012qf}}} \\\hline
& $\ M_h$ $\ $ & $f_h$  $\ $ & $\ M_h$ $\ $ & $f_h$  $\ $	& $\ M_h$  $\ $ & $f_h$ $\ $ \\
\hline\hline
$\eta_c\ $ & 2.98 & 0.272 $\ $ & 2.98 & 0.238 $\ $& 2.98 & 0.279(17)$\ $\\
$\eta_b\ $ & 9.38 & 0.501 $\ $ & 9.39 & / & 9.39 & 0.472(4)\phantom{7}$\ $\\	\hline
\end{tabular*}
\end{table}
%lattice QCD (lQCD) \cite{Davies:2010ip, McNeile:2012qf, Donald:2012ga, Colquhoun:2014ica} --
%$f_{\eta_c}=0.279(17)$, $f_{\eta_b}=0.472(4)$, $f_{J/\Psi}=0.286(4)$, $f_\Upsilon=0.459(22)$;

\smallskip

\addtocounter{section}{1}

\noindent\textbf{\arabic{section}.\,Heavy-light Mesons: Masses and Decay Constants}\,---\,We focus first on the properties of mesons formed from a valence  $c$-quark and $\bar q$-quark, $\hat m_q \leq \hat m_c$.  Namely, beginning with our $\eta_c$ solution, we solve the gap and Bethe-Salpeter equations at a range of evenly spaced values of $\hat m_q <\hat m_c$, directly computing the mass and decay constant of the associated bound-state until that value of $\hat m_q = \hat m_q^{\rm cr}$ is reached for which the kernel defined by Eqs.\,\eqref{KDinteraction}\,--\,\eqref{Mhzeta2} is no longer reliable.
For $D_q$-mesons, this occurs before any moving singularity enters the integration contour used in the RL Bethe-Salpeter equation because the heavy-quark parameters connected with Eq.\,\eqref{Dwconstant} are not appropriate for light quarks.  Since the $s$-quark defines a boundary between dominance of emergent and Higgs mass-generating mechanisms \cite{Ding:2015rkn, Ding:2018xwy}, we terminate direct calculations at $m_q^{\rm cr} =0.4\,{\rm GeV} \approx 4 m_s$.  The value of any desired quantity on $m_q < m_q^{\rm cr}$ is then estimated via extrapolation from $m_q > m_q^{\rm cr}$.  The ambiguity in the value of $m_q^{\rm cr}$ is expressed in the uncertainty bands we place on our extrapolations.

\begin{figure}[!t]
\begin{center}
\includegraphics[clip,width=0.9\linewidth]{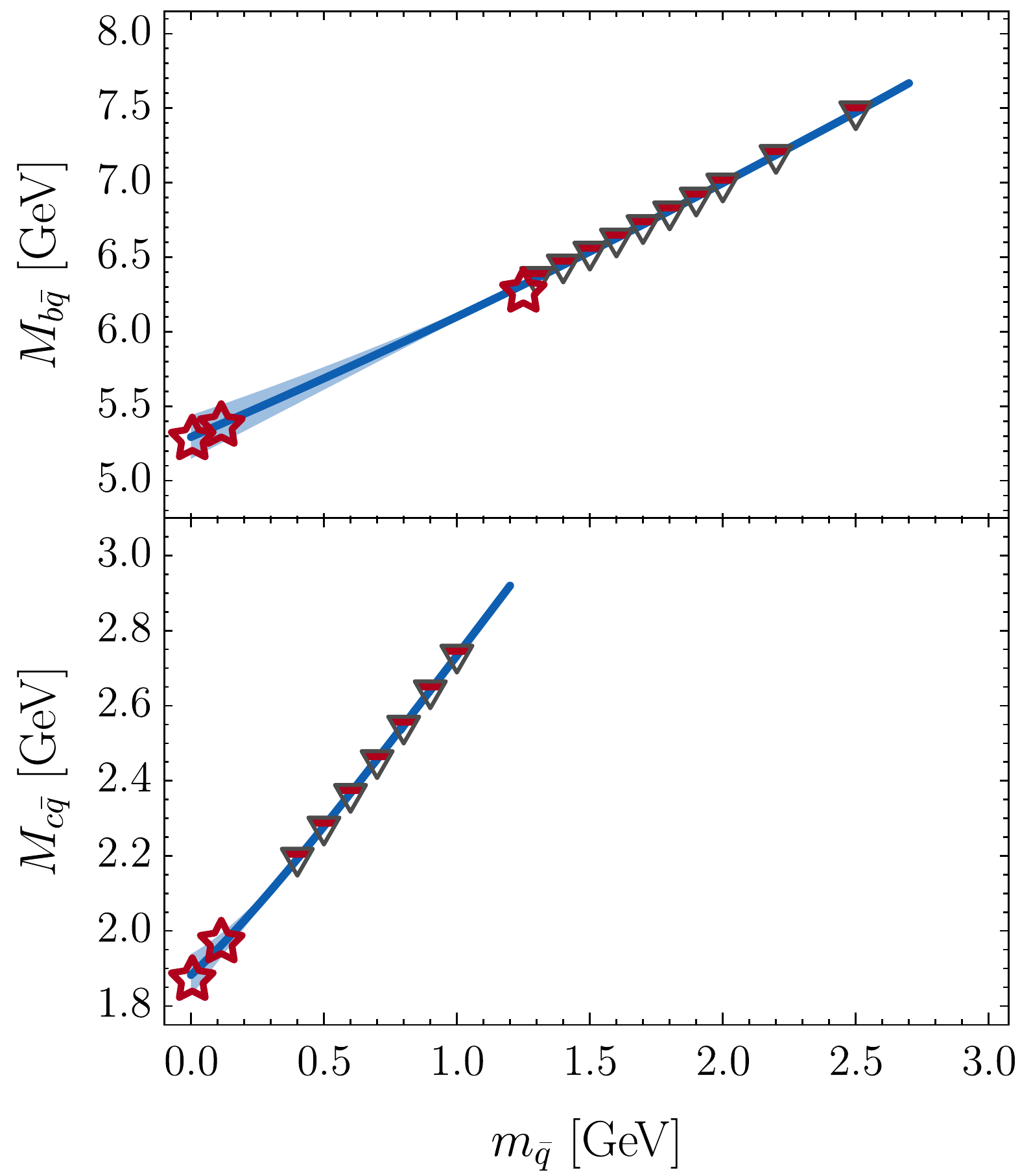}
\end{center}
\caption{\label{Fmasses}  Masses of $c\bar{q}$ (lower panel) and $b\bar{q}$ (upper panel) pseudoscalar mesons computed as a function of the mass of the lighter quark, $m_{\bar q}$.  Triangles -- our computed results and basis for extrapolations (solid blue curves within bands); and open stars -- empirical values listed in Table~\ref{TableQqproperties}.
}
\end{figure}

%%If zeta = 2 GeV were in the one-loop evolution domain, then these masses would correspond to
%%                m_u = (\hat m_u / \hat m_c) * M_c(2) = 0.0049
%%                m_s = (\hat m_s / \hat m_c) * M_c(2) = 0.114
%% Please recomputed all values of light-quark quantities at values of your mass parameters mu=0.0049, ms=0.114.  This will introduce minor changes in the quoted values of extrapolated meson masses, decay constants and PDA parameters.

\begin{table}[!t]
\caption{\label{TableQqproperties}
(\textbf{A}) Masses and decay constants of $D_q$ mesons computed herein, using Eqs.\,\eqref{KDinteraction}\,--\,\eqref{Mhzeta2}; compared with averages of available experimental and lattice-QCD determinations reported in Ref.\,\cite{Tanabashi:2018oca}.
(\textbf{B}) As above for $\bar B_q$ mesons, with lQCD results for $B_c$ taken from Ref.\,\cite{Chiu:2007bc}
(Quantities listed in GeV; and in our normalisation, $f_\pi =0.092\,$GeV.)\vspace*{0.5em}
}
\begin{tabular*}%{l||*{2}{c|}|*{2}{c|}|*{2}{c|}}
{\hsize}
{l@{\extracolsep{0ptplus1fil}}
|c@{\extracolsep{0ptplus1fil}}
c@{\extracolsep{0ptplus1fil}}
|c@{\extracolsep{0ptplus1fil}}
c@{\extracolsep{0ptplus1fil}}
|c@{\extracolsep{0ptplus1fil}}
c@{\extracolsep{0ptplus1fil}}}
\hline
\multicolumn{1}{c}{} & \multicolumn{2}{c}{ Herein } & \multicolumn{2}{c}{ Exp.\,\mbox{\cite{Tanabashi:2018oca}} }& \multicolumn{2}{c}{ lQCD \mbox{\cite{Tanabashi:2018oca}}} \\\hline
(\textbf{A})\;\; & $\ M_h$ $\ $ & $f_h$  $\ $ & $\ M_h$ $\ $ & $f_h$  $\ $	& $\ M_h$  $\ $ & $f_h$ $\ $ \\
\hline\hline
$D_d\ $ & 1.88(5) & 0.158(8) $\ $ & 1.87 & 0.153(7) $\ $ & 1.87 $\ $ & 0.150(1)$\ $\\
$D_s\ $ & 1.94(4) & 0.171(6) $\ $ & 1.97 & 0.177(3) $\ $ & 1.97 $\ $ & 0.176(1)$\ $\\	\hline
\end{tabular*}
\smallskip

\begin{tabular*}%{l||*{2}{c|}|*{2}{c|}|*{2}{c|}}
{\hsize}
{l@{\extracolsep{0ptplus1fil}}
|c@{\extracolsep{0ptplus1fil}}
c@{\extracolsep{0ptplus1fil}}
|c@{\extracolsep{0ptplus1fil}}
c@{\extracolsep{0ptplus1fil}}
|c@{\extracolsep{0ptplus1fil}}
c@{\extracolsep{0ptplus1fil}}}
\hline
\multicolumn{1}{c}{} & \multicolumn{2}{c}{ Herein } & \multicolumn{2}{c}{ Exp.\,\mbox{\cite{Tanabashi:2018oca}} }& \multicolumn{2}{c}{ lQCD \mbox{\cite{Tanabashi:2018oca, Chiu:2007bc}}} \\\hline
(\textbf{B})\;\; & $\ M_h$ $\ $ & $f_h$  $\ $ & $\ M_h$ $\ $ & $f_h$  $\ $	& $\ M_h$  $\ $ & $f_h$ $\ $ \\
\hline\hline
$B_u\,$ & 5.30(15)\, & 0.142(13) $\ $
    & 5.28 \, & 0.138(19) \,
    & 5.28$\phantom{(1)}$\,  & 0.132(3)$\ $\\
$B_s\,$ & 5.38(13)\, & 0.179(12) \,
    & 5.37 \, & / $\ $
    & 5.37$\phantom{(1)}$\,  & 0.161(2)$\ $\\	
$B_c\,$ & 6.31($\phantom{1}1$)\, & 0.367($\phantom{1}1$) $\ $
    & 6.27 \, & / $\ $
    & 6.28(1)\,  & 0.346(3)$\ $\\	\hline
\end{tabular*}
\end{table}
%http://pdg.lbl.gov/2018/mobile/reviews/pdf/rpp2018-rev-pseudoscalar-meson-decay-cons-m.pdf
%http://pdg.lbl.gov/2018/reviews/rpp2018-rev-ckm-matrix.pdf

\begin{figure}[!t]
\begin{center}
\includegraphics[clip,width=0.9\linewidth]{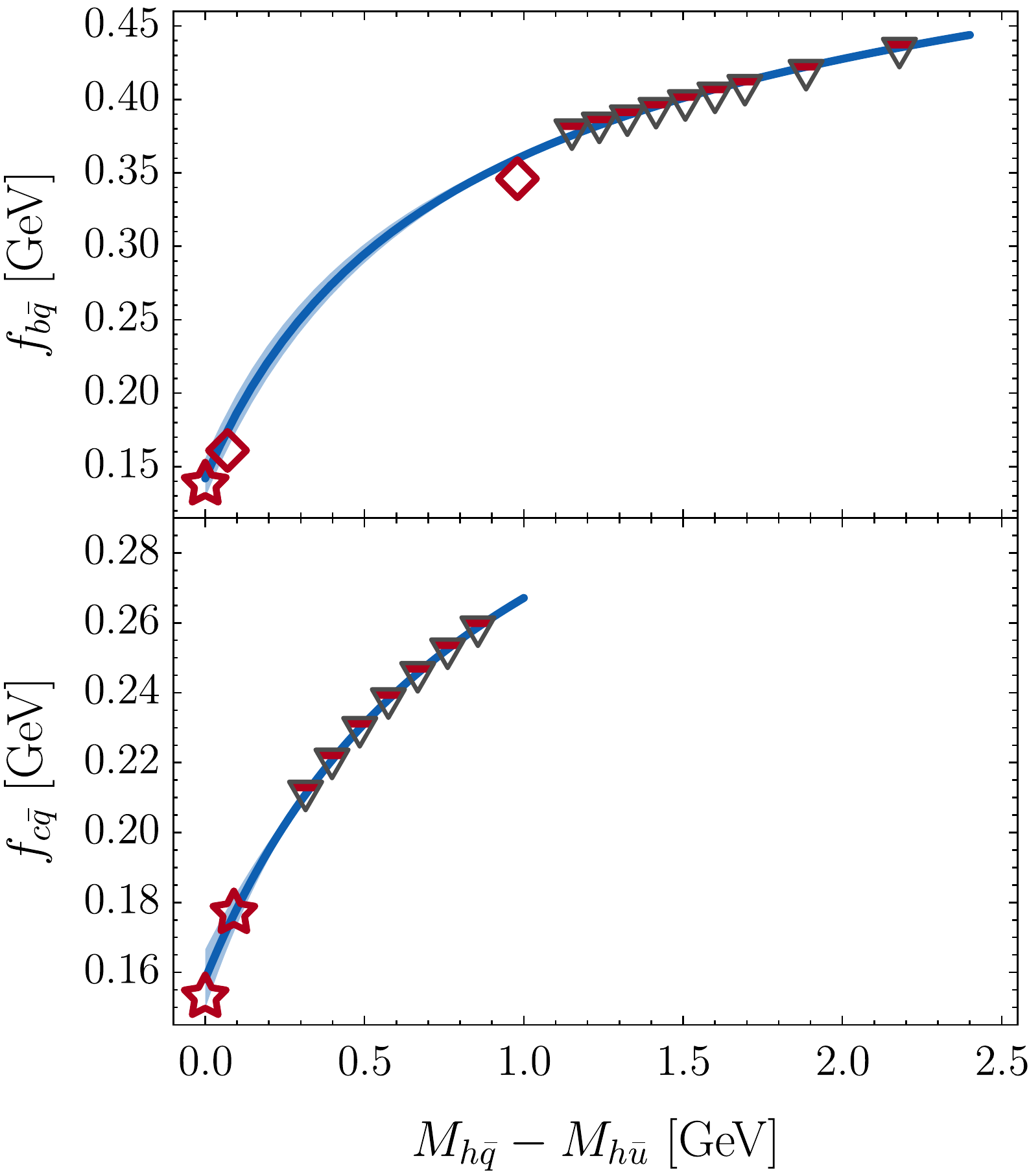}
\end{center}
\caption{\label{Fdecay}  Decay constants of $c\bar{q}$ (lower panel) and $b\bar{q}$ (upper panel) pseudoscalar mesons computed as a function of the mass difference $M_{Q\bar{q}}-M_{Q\bar{u}}$.
Triangles -- our computed results and basis for extrapolations (solid blue curves within bands);
open stars -- empirical values listed in Table~\ref{TableQqproperties};
and open diamonds -- lQCD predictions in Table~\ref{TableQqproperties} (plotted when empirical values are unavailable).
}
\end{figure}

In the lower panel of Fig.\,\ref{Fmasses} we depict the trajectory of $D_q$-meson masses obtained as described above.  Identifying
\begin{equation}
\label{Mhzeta2light}
m_u = M_u(\zeta_2)= 0.0049\,{\rm GeV}\,,\;
m_s = M_s(\zeta_2) = 0.114\,{\rm GeV}\,,
\end{equation}
one therefrom reads the masses in Table~\ref{TableQqproperties}A.  The lower panel of Fig.\,\ref{Fdecay} depicts the associated trajectory of leptonic decay constants, from which one obtains the values listed in Table~\ref{TableQqproperties}A.  Both the masses and decay constants agree well with the empirical values.

We turn now to $\bar B_q$ systems, beginning with our $\eta_b$ solution.  Here a singularity moves into the relevant integration domain for $m_q<m_q^{\rm cr}=1.3\,$GeV, \emph{viz}.\ at a current-mass just above that of the $c$-quark.  The associated trajectory of bound-state masses is depicted in the upper panel of Fig.\,\ref{Fmasses}, from which one extracts the values in Table~\ref{TableQqproperties}B: our predicted $\bar B_q$-meson masses are consistent with experiment.

The upper panel of Fig.\,\ref{Fdecay} displays the mass-dependence of the $\bar B_q$ decay constants.  Since little curvature is evident, it is necessary to introduce the following physical constraints on the extrapolation.
(\emph{i}) Continuum \cite{Ivanov:1998ms} and lQCD \cite{Tanabashi:2018oca} bound-state analyses indicate $f_{B_u} \approx 0.85\,f_{D_{d}}$.  Hence, we require that $f_{B_u}$ take a value in the range $(0.85 \sim 1.0)\,f_{D_{d}}$.
% (ii)
(\emph{ii}) Experiment and available calculations \cite{Chen:2018rwz, Tanabashi:2018oca} indicate that $(f_{Q\bar s} - f_{Q\bar u})/((f_{Q\bar s} + f_{Q\bar u}) \approx 0.09$, independent of the mass-average of the associated bound-states.  We use this feature to constrain $f_{\bar B_s}$ via $f_{\bar B_u}$.
Using the procedure just described, we obtain the curves in the upper panel of Fig.\,\ref{Fdecay} and the associated results in Table~\ref{TableQqproperties}B.

\smallskip

\addtocounter{section}{1}

\begin{figure}[!t]
\begin{center}
\includegraphics[clip,width=0.9\linewidth]{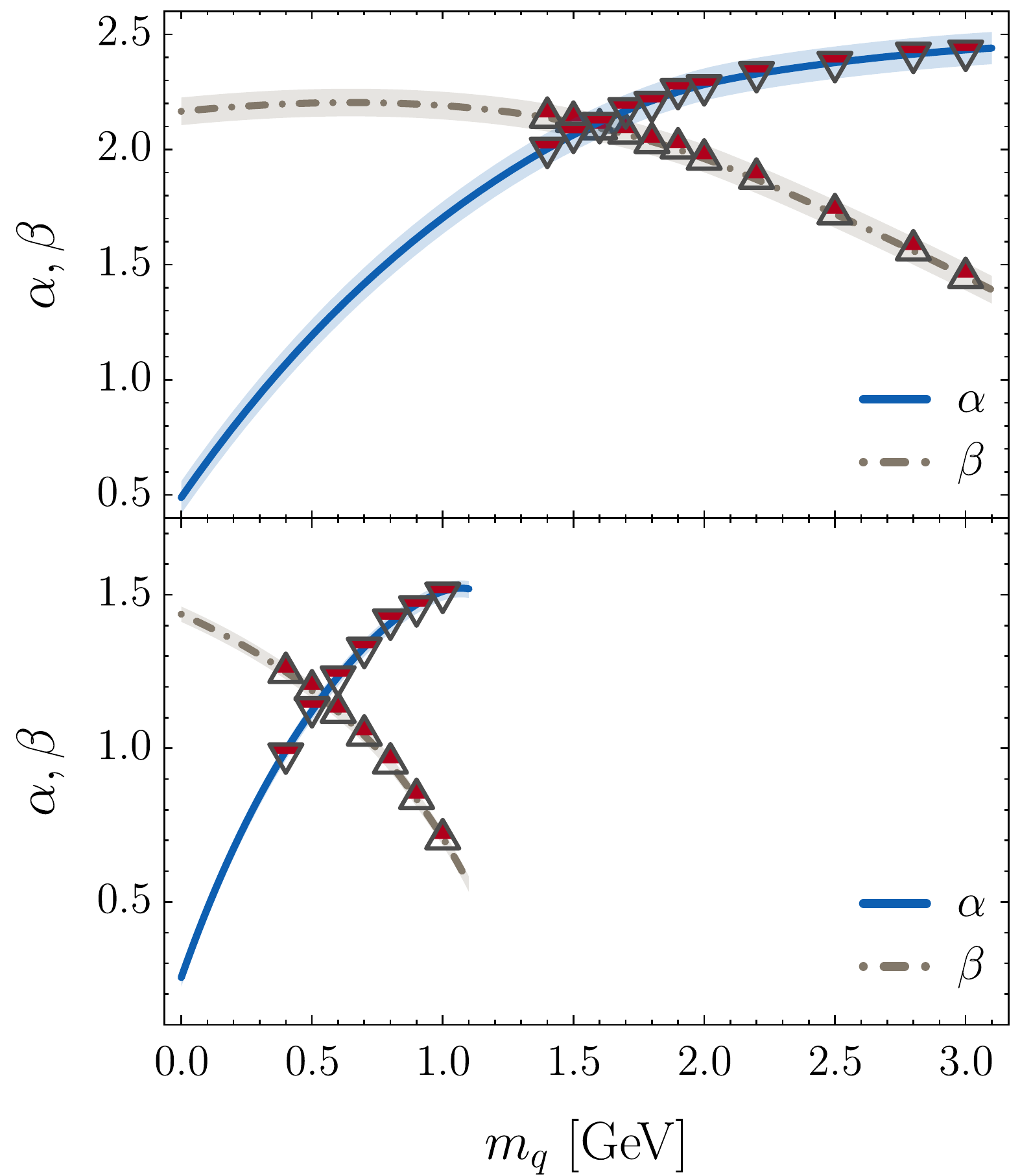}
\end{center}
\caption{\label{Falphabeta}
$(\alpha,\beta)$-pairs specifying, via Eq.\,\eqref{DAfit}, the DAs for $c\bar{q}$ (lower panel) and $b\bar{q}$ (upper panel) pseudoscalar mesons, depicted as a function of the lighter-quark current-mass.
}
\end{figure}

\begin{figure}[!t]
\begin{center}
\includegraphics[clip,width=0.9\linewidth]{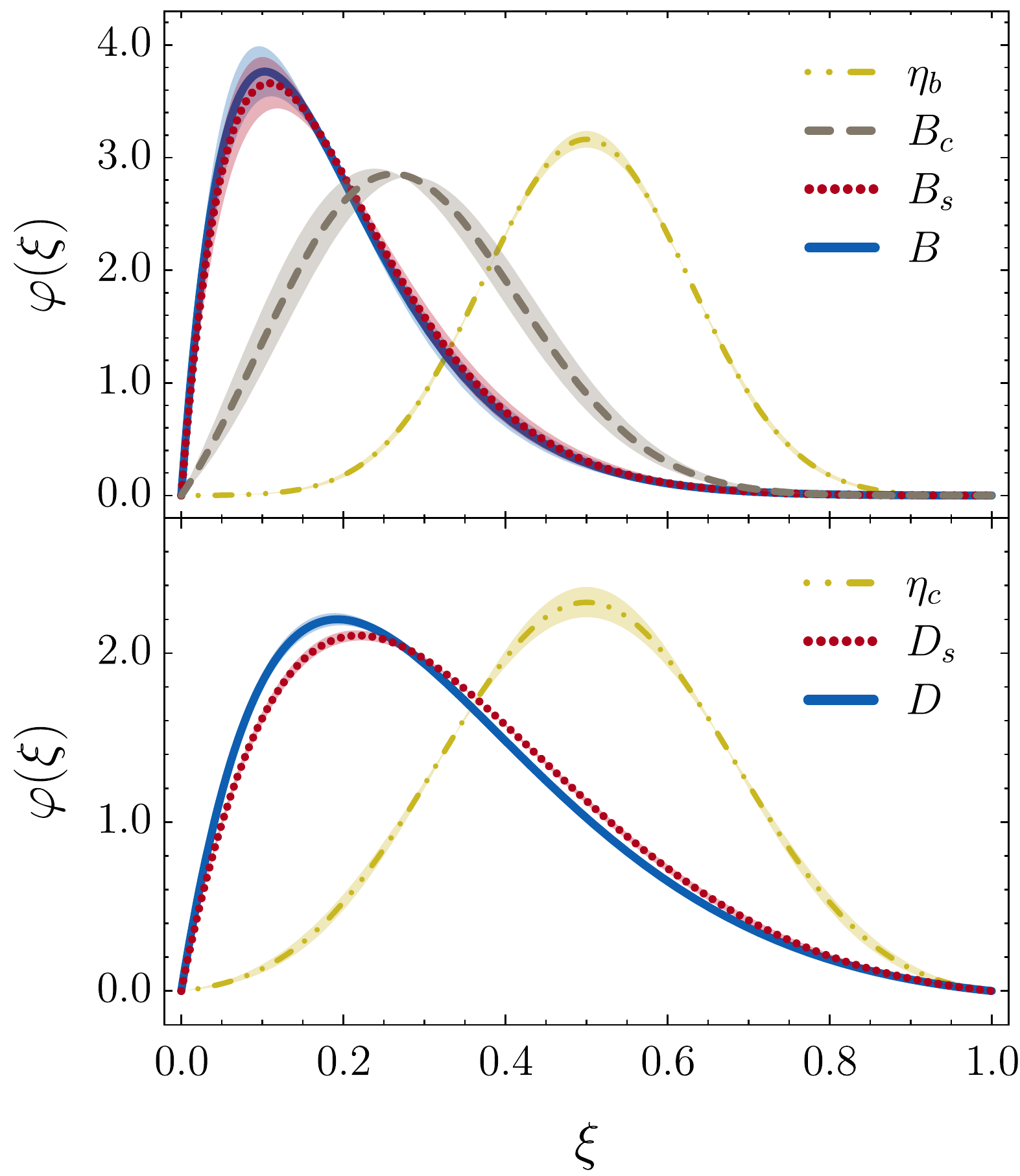}
\end{center}
\caption{\label{FDAs}
Distribution amplitudes of physical heavy-light mesons compared with those of their respective benchmark heavy-heavy systems, computed in the same way \cite{Ding:2015rkn}.
The shaded band surrounding a given curve reflects the uncertainty in the associated values of $(\alpha,\beta)$ listed in Table~\ref{Talphabeta}, which combines that owing to reconstruction from Mellin moments and SPM extrapolation (described in Sec.\,3).
}
\end{figure}

\noindent\textbf{\arabic{section}.\,Heavy-light Mesons: Distribution Amplitudes}\,---\,Re\-turn\-ing to Eqs.\,\eqref{HMDA}, DAs for the systems discussed in the preceding section can be obtained by using the methods introduced in Refs.\,\cite{Chang:2013pq, Ding:2015rkn}.  Namely: (\emph{i}) for each desired and RL-accessible value of the pair $(m_Q,m_{\bar q})$, we compute the Mellin moments
\begin{align}
\langle \xi^m \rangle & := \int_0^1 d\xi \,\xi^m\,\varphi(\xi,\zeta)\,,
\end{align}
$m=1,2,3$; (\emph{ii}) assume that the DA's pointwise form is well represented by\footnote{We have validated this hypothesis by using the maximum entropy method, as described in Ref.\,\cite{Gao:2016jka}, to directly determine the DA in a few, randomly selected cases.}
\begin{align}
\label{DAfit}
\varphi(\xi) = {\mathpzc n}_{\alpha\beta}\,4\xi\bar\xi \, {\rm e}^{-\alpha^2 4 \xi \bar\xi - \beta^2 (\xi-\bar \xi)}\,;
\end{align}
where $\bar \xi = (1-\xi)$ and ${\mathpzc n}_{\alpha\beta}$ ensures $\langle \xi^0 \rangle \equiv 1$;
and (\emph{iii}) determine the coefficient pair $(\alpha,\beta)$ by requiring a least-squares best-fit to $\{\langle \xi^{m=1,2,3}\rangle\}$.  As in Sec.\,4, values of the $(\alpha,\beta)$-pairs relating to systems not directly accessible using RL truncation are then obtained via SPM extrapolation.  Our results for $(\alpha,\beta)$ and their extrapolations are depicted in Fig.\,\ref{Falphabeta}.  The $(\alpha,\beta)$ values for physical mesons are listed in Table~\ref{Talphabeta} and the associated DAs are depicted in Fig.\,\ref{FDAs}.  As anticipated, the DAs become increasingly asymmetric and more sharply peaked as the disparity grows between the current-masses of the meson's valence-quarks.

\begin{table}[t]
\caption{\label{Talphabeta}
$(\alpha,\beta)$-pairs that specify the DAs of heavy-light mesons via Eq.\,\eqref{DAfit}. \mbox{$\,$}
}
\begin{tabular*}%{l||*{2}{c|}|*{2}{c|}|*{2}{c|}}
{\hsize}
{l@{\extracolsep{0ptplus1fil}}
|c@{\extracolsep{0ptplus1fil}}
c@{\extracolsep{0ptplus1fil}}
|c@{\extracolsep{0ptplus1fil}}
c@{\extracolsep{0ptplus1fil}}
c@{\extracolsep{0ptplus1fil}}}
\hline
%\multicolumn{1}{c}{} & \multicolumn{2}{c}{ Herein } & \multicolumn{2}{c}{ Exp.\,\mbox{\cite{Tanabashi:2018oca}} }& \multicolumn{2}{c}{ lQCD \mbox{\cite{Davies:2010ip, McNeile:2012qf}}} \\
 & $D_u$ $\ $ & $D_s$ $\ $ & $\bar B_{u}$ $\ $ & $\bar B_{s}$ $\ $ & $\bar B_{c}$ $\ $ \\\hline\hline
$\alpha\ $ & $0.265(30)\ $ & $0.508(30)\ $ & $0.497(70)\ $ & $0.669(60)\ $ & $1.901(70)\ $ \\
$\beta\ $  & $1.435(30)\ $ & $1.391(30)\ $ & $2.166(60)\ $ &$2.177(60)\ $ & $2.163(60)\ $ \\\hline
\end{tabular*}
\end{table}
%lattice QCD (lQCD) \cite{Davies:2010ip, McNeile:2012qf, Donald:2012ga, Colquhoun:2014ica} --
%$f_{\eta_c}=0.279(17)$, $f_{\eta_b}=0.472(4)$, $f_{J/\Psi}=0.286(4)$, $f_\Upsilon=0.459(22)$;

As in Sec.\,4, values of the $(\alpha,\beta)$-pairs relating to systems not directly accessible using RL truncation are then obtained via SPM extrapolation.  Our results for $(\alpha,\beta)$ and their extrapolations are depicted in Fig.\,\ref{Falphabeta}.  The $(\alpha,\beta)$ values for physical mesons are listed in Table~\ref{Talphabeta} and the associated DAs are depicted in Fig.\,\ref{FDAs}.  As anticipated, the DAs become increasingly asymmetric and more sharply peaked as the disparity grows between the current-masses of the meson's valence-quarks.

With the DAs in hand, it is straightforward to compute a range of moments that play an important role in the application of heavy-quark effective theory (HQET) to exclusive processes involving heavy-light mesons; namely,
\begin{subequations}
\label{HQETMoments}
\begin{align}
\frac{1}{\lambda_h(\zeta)} & = \frac{1}{M_h} \int_0^1d\xi \, \frac{1}{\xi}\,\varphi(\xi;\zeta)\,,\\
\sigma_h(\zeta) & = \frac{\lambda_h(\zeta)}{M_h}
\int_0^1 d\xi\,\frac{\ln[\zeta/(M_h \xi)]}{\xi}
\varphi_h(\xi;\zeta) \,.
\end{align}
\end{subequations}
Another quantity of interest is the mean value of the light-quark light-front momentum within the heavy-light meson:
\begin{equation}
\label{MomFraction}
\langle (w/M_h) \rangle^\zeta := \int_0^1 d\xi\,\xi\,\varphi(\xi;\zeta)\,.
\end{equation}
We list our predictions for these quantities in Table~\ref{Tmoments}.  Notably, $\lambda_h(\zeta)$ decreases with decreasing $\zeta$ \cite{Lee:2005gza}; hence, our computed value of $\lambda_B(\zeta_2)=0.54(3)$ corresponds to $\lambda_B(1\,{\rm GeV}) \approx 0.45(2)$ .

\begin{table}[t]
\caption{\label{Tmoments}
Moments in Eqs.\,\eqref{HQETMoments}, \eqref{MomFraction}, evaluated at $\zeta = \zeta_2 = 2\,$GeV.  For comparison, Ref.\,\cite{Lee:2005gza} reports $\lambda_{\bar B_u}(\zeta_2) = 0.58(4)$, $\sigma_{\bar B_u}(\zeta_2)=1.95(7)$.
\vspace*{0.5ex}
}
\begin{tabular*}%{l||*{2}{c|}|*{2}{c|}|*{2}{c|}}
{\hsize}
{l@{\extracolsep{0ptplus1fil}}
|c@{\extracolsep{0ptplus1fil}}
c@{\extracolsep{0ptplus1fil}}
|c@{\extracolsep{0ptplus1fil}}
c@{\extracolsep{0ptplus1fil}}
|c@{\extracolsep{0ptplus1fil}}}
\hline
%\multicolumn{1}{c}{} & \multicolumn{2}{c}{ Herein } & \multicolumn{2}{c}{ Exp.\,\mbox{\cite{Tanabashi:2018oca}} }& \multicolumn{2}{c}{ lQCD \mbox{\cite{Davies:2010ip, McNeile:2012qf}}} \\
 & $D_u$ $\ $ & $D_s$ $\ $ & $\bar B_{u}$ $\ $ & $\bar B_{s}$ $\ $ & $\bar B_{c}$ $\ $ \\\hline\hline
%
%$\lambda_h(\zeta)\ $ & $0.329(9)\ $ & $0.368(9)\ $ & $0.539(29)\ $ & $0.557(27)\ $ & $1.301(70)\ $ \\
%$\sigma_h(\zeta)\ $  & $2.340(11)\ $ & $2.204(9)\ $ & $1.891(158)\ $ &$1.838(155)\ $ & $0.758(119)\ $ \\
%$\langle (w/M_h) \rangle^\zeta\ $  & $0.317(4)\ $ & $0.333(4)\ $ & $0.194(7)\ $ &$0.197(8)\ $ & $0.293(10)\
$\lambda_h(\zeta)/{\rm GeV}\ $ & $0.33(1)\ $ & $0.37(1)\ $ & $0.54(\phantom{6}3)\ $ & $0.56(\phantom{6}4)\ $ & $1.30(\phantom{6}7)\ $ \\
$\sigma_h(\zeta)\ $  & $2.34(1)\ $ & $2.20(1)\ $ & $1.89(16)\ $ &$1.84(16)\ $ & $0.76(12)\ $ \\
$\langle (w/M_h) \rangle^\zeta\ $  & $0.32(1)\ $ & $0.33(1)\ $ & $0.19(\phantom{6}1)\ $ &$0.20(\phantom{6}1)\ $ & $0.29(\phantom{6}1)\ $ \\\hline
$2 \lambda_h(\zeta)/M_h$ & $0.35(1)\ $& $0.38(1)\ $ & $0.20(\phantom{6}1)\ $ & $0.21(\phantom{6}1)\ $ & $0.44(\phantom{6}4)\ $\\\hline
\end{tabular*}
\end{table}

It is interesting to note that if one were to assume $\varphi_h(w) \approx \varphi_h^{\rm e}(w) = (w/\lambda_h^2) \exp(-w/\lambda_h)$, then $\langle (w/M_h) \rangle = 2 \lambda_h/M_h$.  We have entered these values as Row~4 in Table~\ref{Tmoments}.  Evidently, by this measure, $\varphi_h\approx \varphi_h^{\rm e}$ provides a fair approximation for heavy-light systems.

\begin{figure}[!t]
\begin{center}
\includegraphics[clip,width=0.9\linewidth]{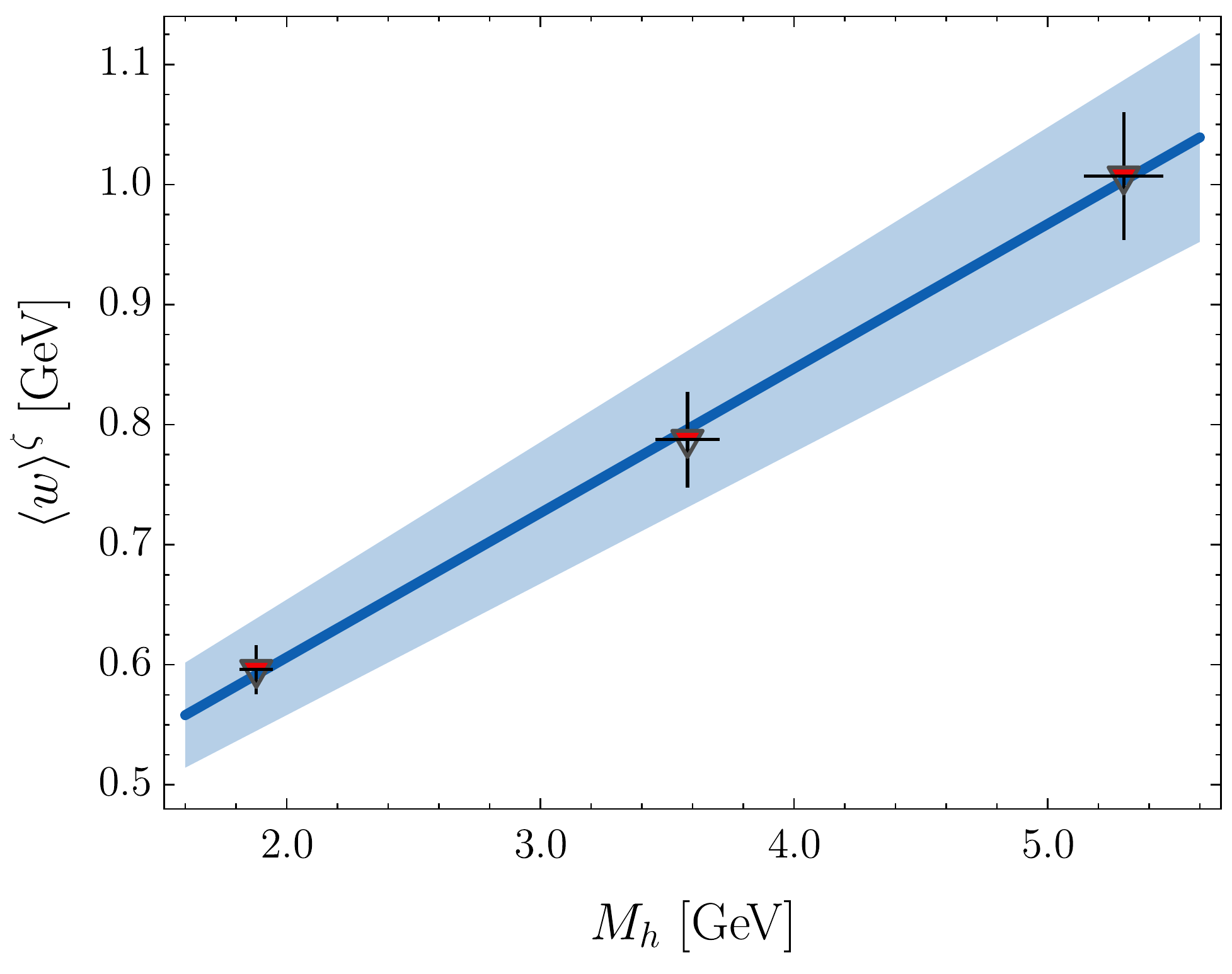}
\end{center}
\caption{\label{FwM}
Mean light-front momentum of a light-quark with zero current-mass as a function of the current-mass of its partner in the heavy-light system, with the latter measured by the bound-state's mass.  The solid (blue) line is the curve in Eq.\,\eqref{limitingxi}.  The blue band provides a conservative indication of the uncertainty introduced by that on each of the inputs.
}
\end{figure}

We have also computed $\langle w \rangle^\zeta$ at $m_Q=m_c, (m_c+m_b)/2, m_b$ in the limit $m_{\bar q}\to 0$, with the results depicted in Fig.\,\ref{FwM}.  They are described by a straight line, which translates into the following behaviour:
%%c0 = 0.365554 +/- 0.0374068 c1 = 0.120299 +/- 0.0152948
\begin{subequations}
\label{limitingxi}
\begin{align}
\langle \xi \rangle^{\zeta_2} &= \langle (w/M_h) \rangle^{\zeta_2}=  \xi_0^{\zeta_2} + \frac{\xi_1^{\zeta_2}}{M_h}\,,\\
\xi_0^{\zeta_2} &= 0.120(11)\,,\;
\xi_1^{\zeta_2} = 0.366(26)\,{\rm GeV}.
\end{align}
\end{subequations}
This result and related algebraic analysis using the methods of Refs.\,\cite{Ivanov:1997yg, Ivanov:1998ms} indicate that for each value of $\zeta$, $\langle \xi \rangle^{\zeta}\to\xi_0^\zeta$, \emph{i.e}.\ the light-quark light-front momentum-fraction takes a finite, nonzero value in the limit $M_h\to \infty$.  Naturally, at any large, fixed value of $M_h$, $\xi_0^{\zeta}\to1/2$ as $\zeta \to\infty$.

\begin{table}[t]
\caption{\label{TBF}
Branching fraction for the $B\to \gamma(E_\gamma) \ell \nu_\ell$ radiative decay as a function of the minimum photon energy, $E_\gamma^{\rm min}$.
\underline{Row~1}.\ Computed using our predictions for $m_b$, $M_B$, $f_B$, $\lambda_B$: Eqs.\,\eqref{Mhzeta2}, \eqref{Mhzeta2light} and Tables~\ref{TableQqproperties}, \ref{Tmoments}.
\underline{Row~2}.\ Computed using $m_b$, $M_b$, $f_B$ from elsewhere \cite{Tanabashi:2018oca}.
Comparable entries agree within errors.
Contemporary data indicate $\Gamma_{B\to \gamma \ell \nu_\ell}/\Gamma_B < 3.0 \times 10^{-6}$ for $E_\gamma^{\rm min} = 1\,$GeV \cite{Gelb:2018end}.
%% Experiment doesn't need Vub to extract this ... they just measure the ratio.
%
($\Gamma_B = 0.401(1)\,$meV \cite{Tanabashi:2018oca}.)
\vspace*{0.5em}
}
\begin{tabular*}%{l||*{2}{c|}|*{2}{c|}|*{2}{c|}}
{\hsize}
{l@{\extracolsep{0ptplus1fil}}
|c@{\extracolsep{0ptplus1fil}}
c@{\extracolsep{0ptplus1fil}}
c@{\extracolsep{0ptplus1fil}}}
\hline
%\multicolumn{1}{c}{} & \multicolumn{2}{c}{ Herein } & \multicolumn{2}{c}{ Exp.\,\mbox{\cite{Tanabashi:2018oca}} }& \multicolumn{2}{c}{ lQCD \mbox{\cite{Davies:2010ip, McNeile:2012qf}}} \\
$E_{\gamma}^{\rm min}/{\rm GeV}\ $ & $1.0\ $ & $1.5\ $ & $2.0\ $ \\\hline
\multirow{2}{*}{$10^6\, \Gamma_{B\to \gamma \ell \nu_\ell}/\Gamma_B\ $}
& $0.84(25)\ $ & $0.47(15)\ $ & $0.17(6)\ $\\
&  $0.77(24)\ $ & $0.43(14)\ $ & $0.15(5)\ $\\\hline
\end{tabular*}
\end{table}
% errors
% dlB = +/- 0.09
% dfB = 0.13
% dcv = 0.056
% dmb = 0.11
% dVub =  0.13
% dR = 0.06
%% Radiative corrections ... NLL ... R(E,zeta)/lambda_B(zeta) = constant on hard-collinear-scale \in [1,2] ... this means I should be able to use our zeta = 2 value with the plotted function.

\begin{figure}[!t]
\begin{center}
\includegraphics[clip,width=0.9\linewidth]{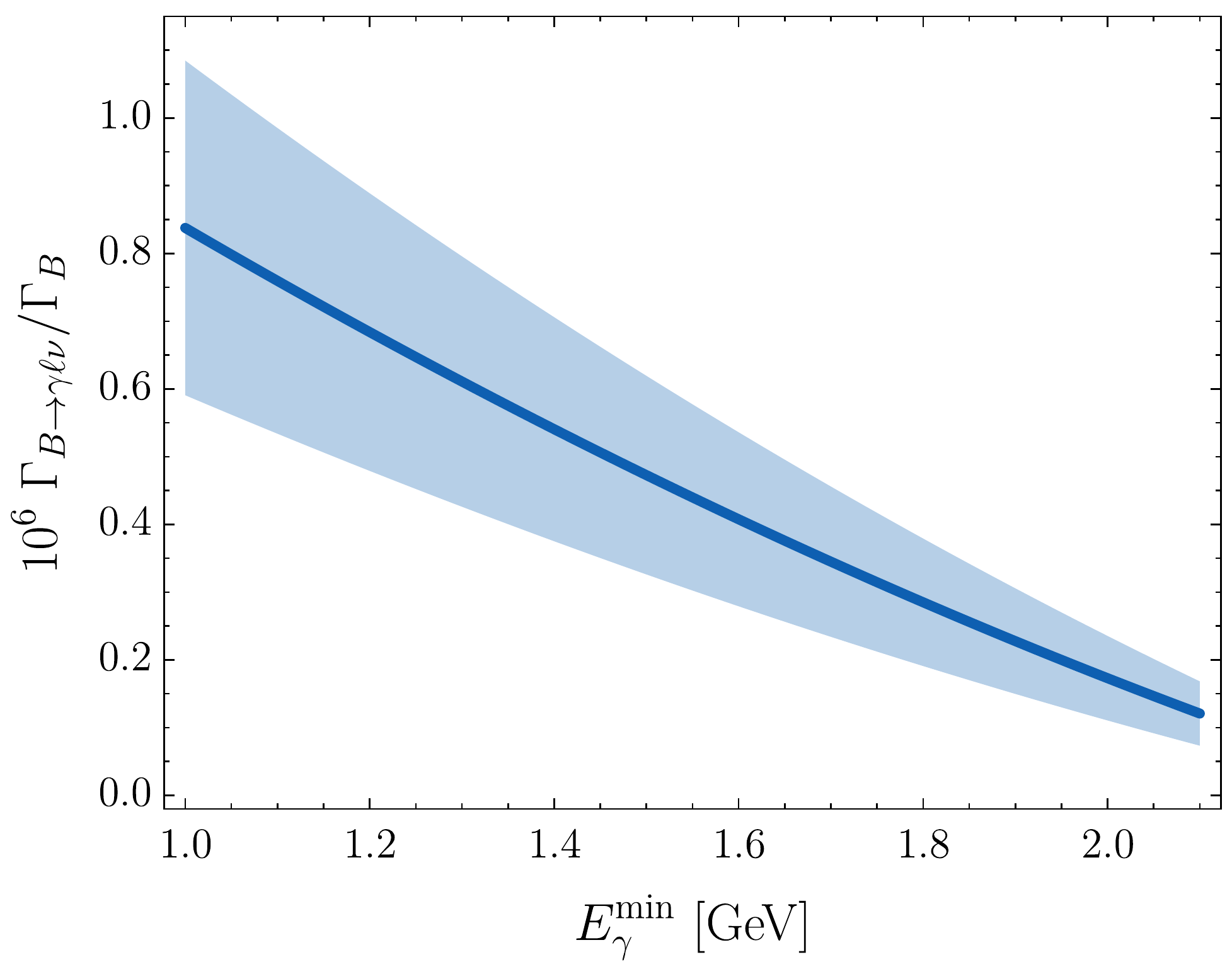}
\end{center}
\caption{\label{Branching}
$E_\gamma^{\rm min}$-dependence of the $B\to \gamma \ell \nu_\ell$ branching fraction computed using our predictions for $m_b$, $M_B$, $f_B$, $\lambda_B$: Eqs.\,\eqref{Mhzeta2}, \eqref{Mhzeta2light} and Tables~\ref{TableQqproperties}, \ref{Tmoments}: the blue band provides a conservative indication of the uncertainty introduced by that on each of the inputs.
}
\end{figure}

%% R(E,\mu) is plotted in Beneke:2011nf ... it is basically flat at NLL as a function of \zeta
%% Also plot energy dependence
We now follow Refs.\,\cite{Beneke:2011nf, Braun:2012kp, Beneke:2018wjp} and compute the branching fraction for the $B\to \gamma \ell \nu_\ell$ radiative decay.  This process is analogous to the $\gamma^\ast \gamma \to \pi^0$ transition in the sense that it is amenable to analysis using the factorisation formalism, depends linearly upon the participating meson's DA, and is the simplest process to probe that DA.
In this calculation, we employ the formula for the $E_\gamma$-dependent differential decay width in Refs.\,\cite{Beneke:2011nf, Braun:2012kp}, which retains $1/m_B^2$ and $1/E_\gamma^2$ corrections, but our predictions for $m_b$, $M_B$, $f_B$, $\lambda_B$: Eqs.\,\eqref{Mhzeta2}, \eqref{Mhzeta2light} and Tables~\ref{TableQqproperties}, \ref{Tmoments}.
Assuming that the factorised expression is valid for $E_\gamma > E_\gamma^{\rm min}$, we integrate over $E_\gamma \in [E_\gamma^{\rm min},E_\gamma^{\rm max}=m_B/2]$ to obtain the branching fractions in Table~\ref{TBF} when $|V_{ub}| = 3.94(36) \times 10^{-3}$ \cite{Tanabashi:2018oca}.
%%\begin{equation}
%%\begin{array}{l|ccc}
%%E_{\gamma}^{\rm min}/{\rm GeV} & 1.0 & 1.5 & 2.0 \\\hline
%%10^6\, \Gamma_{B\to \gamma \ell \nu_\ell}/\Gamma_B &
%%            0.84(25) & 0.47(15) & 0.17(6)
%%\end{array}\,,
%%\end{equation}
%where $\Gamma_B = 0.401(1)\,$meV and we used $|V_{ub}| = 3.94(36) \times 10^{-3}$ \cite{Tanabashi:2018oca}.
Our computed $E_\gamma^{\rm min}$ dependence of the branching fraction is depicted in Fig.\,\ref{Branching}. At present, for a fixed value of $\lambda_B$, the largest sources of error are $|V_{ub}|$ and $f_B$, which appear quadratically in the numerator of the differential decay-width formula.
Notably, if we choose to artificially change $\lambda_B \to \tfrac{2}{3} \lambda_B$, the computed values of the branching fraction become approximately $2.6$-times larger.  Such marked sensitivity to the $B$-meson DA has previously been highlighted \cite{Beneke:2011nf, Braun:2012kp}.

\smallskip

\addtocounter{section}{1}

\noindent{\textbf{\arabic{section}.\,Epilogue}}\,---\,%
Working with the leading-order, symmetry-preserving truncation of the relevant Dyson-Schwinger equations and an interaction kernel constrained by analyses of QCD's gauge sector and tested in studies of heavy-heavy mesons and triply-heavy baryons, we delivered parameter-free predictions for the masses, decay constants and light-front distribution amplitudes of heavy-light mesons.
No material betterment of these results can be anticipated before either sound improvements over the leading-order truncation of the continuum bound-state problem have been developed for heavy-light systems or numerical simulations of lattice-regularised QCD become capable of simultaneously computing all these quantities at physical current-quark masses on large lattices with small interstitial spacing.

Owing to its importance as a basic test of the factorisation approach to hard exclusive processes in QCD, we used our results to calculate the branching fraction for the radiative decay $B\to \gamma \ell \nu_\ell$.  Precision measurements at new-generation $B$-factories can test this prediction and, hence, bring within reach an empirical check on the validity of factorisation in the treatment of exclusive decays of heavy-light mesons
%Hence, an empirical check on the validity of factorisation in the treatment of exclusive decays of heavy-light mesons is within reach.

\smallskip

%\section*{Acknowledgments}
%\noindent\textbf{Acknowledgments}.
%
\noindent\textbf{Acknowledgments}\,---\,%We are grateful for insightful comments from .
%\acknowledgments
%
%
We are grateful to:
V.~Braun for constructive comments made during the \emph{Workshop on Mapping Parton Distribution Amplitudes and Functions}, September 2018, European Centre for Theoretical Studies in Nuclear Physics and Related Areas (ECT*), Trento, Italy;
ECT* and its resources during that and the following \emph{Workshop on Emergent mass and its consequences in the Standard Model};
the University of Huelva, Huelva - Spain, and the University of Pablo de Olavide, Seville - Spain, for their hospitality and support during the \emph{4th Workshop on Nonperturbative QCD} at the University of Pablo de Olavide, November 2018.
Work supported by:
the Chinese Government's \emph{Thousand Talents Plan for Young Professionals};
Spanish MEyC, under grants FPA2017-84543-P and SEV-2014-0398;
Chinese Ministry of Education, under the \emph{International Distinguished Professor} programme;
Jiangsu Province \emph{Hundred Talents Plan for Professionals};
and U.S.\ Department of Energy, Office of Science, Office of Nuclear Physics, under contract no.\,DE-AC02-06CH11357.

\centerline{\rule{0.8\linewidth}{0.1ex}}

%\bibliographystyle{../../../h-physrev4}
%\bibliography{../../../CollectedBiB}
%%\bibliographystyle{../../../../zProc/z10/z10KITPC/h-physrev4}
%%\bibliography{../../../../CollectedBiB}
%%\bibliographystyle{../../zchanglei/zchanglei2/zPDA5/model1a-num-names}
%%\bibliography{../../../CollectedBiB}

\end{document}